\documentclass[12pt]{article}
\usepackage{mathtext}
\usepackage{graphics}
\usepackage{float}
\restylefloat{figure}
\usepackage{subfigure}
\usepackage{epsfig}
\title{\bf HELIOGRAPH OF THE UTR-2 RADIO TELESCOPE}
\author{\bf A. A. Stanislavsky\footnote{e-mail: alexstan@ri.kharkov.ua}, A. A. Koval, A. A. Konovalenko\\ \bf and \framebox{E. P.
Abranin}\\ \\
{\it Institute of Radio Astronomy, 4 Chervonopraporna St.,}\\
{\bf\it 61002 Kharkiv, Ukraine}}

\begin{document}
\large\tolerance8000\hbadness10000\emergencystretch3mm\maketitle
\thispagestyle{empty}

\begin{abstract}
The broadband analog-digital heliograph based on the UTR-2 radio
telescope is described in detail. This device operates by
employing the parallel-series principle when five equi-spaced
array pattern beams which scan the given radio source (e.g. solar
corona) are simultaneously shaped. As a result, the obtained image
presents a frame of 5 $\times$ 8 pixels with the space resolution
25$\,'\times$ 25$\,'$ at 25 MHz. Each pixel corresponds to the
signal from the appropriate pattern beam. The most essential
heliograph component is its phase shift module for fast sky
scanning by pencil-shape antenna beams. Its design, as well as its
switched cable lengths calculation procedure, are presented, too.
Each heliogram is formed in the actual heliograph just by using
this phase shifter.  Every pixel of a signal received from the
corresponding antenna pattern beam is the cross-correlation
dynamic spectrum (time-frequency-intensity) measured in real time
with the digital spectrum processor. This new generation
heliograph gives the solar corona images in the frequency range
8-32 MHz with the frequency resolution 4 kHz, time resolution to 1
ms, and dynamic range about 90 dB. The heliographic observations
of radio sources and solar corona made in summer of 2010 are
demonstrated as examples.
\end{abstract}

\section{Introduction}
Comprehensive information on the physical processes that accompany
solar activity can be obtained only with engaging of an extensive
scope of observation means. At the meter wavelengths, radio
spectrographs, heliographs and polarimeters are widely used to
meet this purpose \cite{kruger}. A worse situation is the case for
decameter wavelengths. In particular, the heliograph realization
is technically a rather labor intensive problem. Obtaining the
high-resolution images (being one of the major demands in
application of heliographs) at that low frequencies requires the
huge size antennas ($>$ 1 km). At the same time, the heliograph
can help obtain new useful information on the extremely
multifarious types of bursts being rather specific to this range
of frequencies.

Any heliograph has essential advantages in the case it ensures the
possibility of simultaneous observations at several frequencies.
Multi-frequency measurements of the positions of sources of Type
II and III bursts allow to immediately determine the velocity of
shock waves and this of electron beams which are propagated in the
solar corona. Of substantial interest are the measurements of the
velocity of electron beams generating Type III and IIIb bursts at
different trajectory phase. The positional measurements at several
frequencies will probably permit evaluation of the magnitude of
visible displacement of burst sources with respect to their true
position caused by the refraction and scattering of radio waves in
corona.

Thus, the heliograph permits observation of an angular structure
of burst sources and its evolution during their lifetime, as well
as to identify the bursts with the appropriate activity regions
and measure their heights above the photosphere. The position
measurements of double Type IIIb-III, III-III bursts and drifting
pairs allow to find out whether the both components are excited in
the same place of the corona and at which plasma frequency. Of
special interest are the investigations of properties and
determination of the positions of coronal mass ejections,
frequently associated with the Type II bursts. This phenomenon is
considered a key one in the problems of solar-terrestrial
relationships and space weather.

No less interesting direction of the research efforts are
observations of the two-dimensional brightness distributions
across the quiet Sun at the decameter waves. At the same time, as
is known, just at the heights where the decameter radiation is
generated there exists the most interesting corona region where
the solar wind originates. The here-presented far from exhaustive
list of problems capable of solving with the heliograph makes its
construction a rather attractive idea. Especially, in case if
there is an opportunity of using the available antenna system
suitable for this purpose.

To date, there exist not too many radio telescopes permitting to
obtain 2D heliograms of the Sun and other sky radiation sources.
In the microwave range, they are represented by the Nobeyama
heliograph (Japan) operating at two frequencies, 17 and 35 GHz
\cite{Nishio98}, and the RATAN-600 based heliograph (Russia)
operating at 3.75 GHz \cite{Opeikina95}. Roughly within these
frequency ranges, the observations are also made at the Siberian
Solar Radio Telescope (Russia) \cite{Grechnev03} at 5.7 GHz, and
at the Owens Valley Radio Observatory interferometer (USA) within
1-18 GHz, whose whole range being split into 86 individual
subbands \cite{Padin91}. The anticipated Chinese heliograph being
designed for radio measurements within 0.4-15 GHz \cite{Yi-Hua06}
can be considered belonging under the promising instruments, too.
For centimeter and meter wavelength radio astronomy investigation,
used are the Nancay radioheliograph (France) \cite{NRHG93},
operating only at frequencies 169, 327 and 408 MHz, and the
Gauribidanur radioheliograph (India) which surveys at some
individual frequencies within 40-150 MHz range \cite{Ramesh98}.
Since 1969 till 1984, the Culgoora radioheliograph (Australia)
\cite{Sheridan83} was functioning at frequencies 43, 80, 160 and
327 MHz. Note that very few instruments were used at the decameter
wavelengths. In this respect we may mention the Clark Lake
Teepee-Tee radio telescope (USA) \cite{Erickson82} at which the
measurements at some individual frequencies within  15-125 MHz
were made. Though for now, it is already put out of service. It
will be observed also that developing the antenna arrays capable
of operating low frequencies (decameter wavelengths) is connected
with heavy technological and methodological difficulties. This
frequency range is subject to very heavy noise conditions, the
ionosphere exerts an essential impact on the behavior of decameter
wave propagation, and achieving the narrow beam pattern requires
antennas of huge areas. For instance, the Nancay Decameter Array
(Nancay, France) \cite{Lecacheux2000} is effectively used for
getting dynamic spectra of solar radio emission, while because of
its small sizes practically never used for heliographic
observations. On account of these reasons, nowadays only a few
decameter range antenna systems exist capable of performing
heliographic observations, though they also have limited
capacities, and thus need upgrading, or even new antenna systems
should be built. Suffice it to mention the promising projects
LOFAR \cite{white,Reich07} and LWA \cite{Kassim05} which are to be
put into service in the near future. Therefore the researches
carried out with the UTR-2 based heliograph are of great
scientific interest, making up for a deficiency in the knowledge
of physical processes on the Sun and/or in the circumsolar space.
Note that until now, the UTR-2 radio telescope remains the
world-largest and most effective instrument operating within 8-32
MHz, and it is expected to remain that kind performer in long-term
future. Accordingly, its intensive design improvement,
instrumentation and usage (this including in heliographic mode,
too) promises a vast deal of useful astrophysical information.

The high solar activity is frequently accompanied by sporadic
radio emission in the meter and decameter wavelength ranges
\cite{kruger,pick2008}. Some of its types (I, II, III and IV) are
common for these ranges. At the same time, they have their
inherent specificity. For instance, the Type I  bursts, most
frequent at meter waves, are practically not observed at the
decameter wavelengths. On the other hand, the short-lived
stria-bursts, drifting pairs, Type IIIb bursts are met only at
frequencies below 60 MHz, though the latter, along with Type III
bursts, are the most numerous events of the solar decameter
radiation \cite{Abranin78}.

The solar decameter sporadic radiation has begun to be
systematically studied with the work of Ellis and McCulloch
\cite{Ellis67}. Since then, a substantial progress was reached in
understanding its nature. Meanwhile, the theoretical formulation
of the mechanism and conditions of solar burst generation cannot
yet give, in most cases, an adequate representation. Therefore
further analysis of new experimental data, including those
observed with the heliograph, is of current importance.

The present paper is devoted to the UTR-2 heliograph construction,
its general functional scheme and discussion of its most important
features. In Section~\ref{par2} we start with the history of
heliograph studies in our institute. The basis of any
radioheliograph, the UTR-2 based heliograph making no exception,
is the antenna system. In Section~\ref{par3}, this latter brief
description as applied to the heliograph design, as well as the
detailed analysis of the modern configuration of a two-dimensional
decameter wavelength heliograph based on the UTR-2 T-shaped
antenna system (in Sections~\ref{par4} and \ref{par5}), are
presented. Next, Section~\ref{par6} is aimed at design description
of the most important UTR-2 based heliograph element, namely the
phase shifter for fast beam scanning. The phase shifter essential
parts are the switched coaxial delay-line cables. In
Section~\ref{par7} the method is described for calculating the
lengths of these switched cables. The heliograph control unit
helps to generate the pulse train for switching the fast scanning
phase shifter and scan markers generation (see
Section~\ref{par9}). Consequently, the UTR-2 array pattern beam is
scanning the desired sky area, and the heliograph scan sector
format is discussed in Section~\ref{par8}. Further, in
Section~\ref{par10} we consider the operation features of the
heliograph receiver-recorder. It is based on a multichannel
digital spectropolarimeter (DSP). In Section~\ref{par11} we
present the preliminary results of test observations made with the
considered instrument in summer of 2010. In conclusions, we
summarize the heliograph development and prospects for its future
application in radio astronomy observations at decameter
wavelengths.

\section[Brief History of Heliograph Studies in the UTR-2]
{Brief History of Heliograph Studies in the UTR-2
Observatory}\label{par2} The first positional observations of
solar radiation with the use of a two-dimensional heliograph based
on the UTR-2 radio telescope were made during July 31 to August
11, 1976 \cite{Abranin80}. At that time, the enhanced radiation
was associated with emerging of the active McMath No. 14352 region
on the eastern limb which, while moving westwards, was
intersecting the entire solar disk. For the aforesaid period of
observations, numerous Type III and IIIb bursts, Type IIIb-III
bursts, as well as short-lived narrow-band stria-bursts being
components of the Type IIIb burst chains were recorded. The daily
two-hour measurements centered for the local midday were made at
the operating frequency 25 MHz, and also in parallel by the
``North-South'' (N-S) antenna array to which output the dynamic
spectrograph for the 23.5-25.5 MHz band was mounted. The recorded
heliograms permitted constructing the histograms of coordinate
distribution of effective centers of Type III and stria-bursts for
hour angle ($h$) and declination ($\delta$) in the selected days
for different positions of solar disk active regions. The
positional data on the average coordinates of burst sources
centers were used to obtain the day-to-day dependence of the
radiating region position. The motion velocity of the active
region, crossing the solar disk westwards, permitted estimates of
the average heights $R_s$ (being radial distances from the Sun's
center) of the Type III burst and stria-burst sources during the
radio burst. The heights $R_s$ were determined by several methods
which yielded close results. In the meter and centimeter
wave-lengths, the radial distances of the coronal emitting regions
were calculated in papers \cite{Malinge60,Clavelier68}. The $R_s$
values estimated with the UTR-2 heliographic observations have
appeared comparable to each other and corresponded to those first
obtained at decameters during the interference measurements at the
Clark Lake Radio Observatory (CA, USA) at 30 MHz while
investigating a number of solar noise storms \cite{Gergely75}. The
heliograms have permitted the conclusion that the stria-bursts of
IIIb chains and Type III bursts appear about the same place on the
Sun's image plane.

Further research efforts, assisted by the UTR-2 antenna system
operating the mode of a two-dimensional heliograph, were largely
concentrated on studying the Type IIId radiobursts with echo
components
\cite{Abranin82,Abranin84,Abranin96,Abranin97,Abranin98,Abranin99,Abranin00}.
The decameter Type IIId bursts were first recorded at the UTR-2
observations of July 6, 1973. Narrow-band elements of this
emission fine structure -- the diffuse stria-bursts -- are a
variety of ordinary short-lived stria-bursts, of which the dynamic
spectra of Type IIIb bursts are patterned. The principal feature
of diffuse stria-bursts consists in that their form largely
depends on heliolongitude of the coronal sources region. When the
Type IIId source is in the near-limb region, the stria-bursts with
steep fronts and sharp peaks are observed monotonically damping in
a few seconds. With the active region approach to the central
solar meridian, the time splitting occurs, i.e. the echo-like
component of bursts with growing delay appears. The echo-burst
delay time becomes maximum when the emitting region traverses the
central meridian. The appearance of an extra burst echo-component
and the delay variation with heliolongitude are qualitatively
explained by using a simple model of a point pulsed source placed
into a spherically symmetric corona and emitting at the plasma
frequency second harmonic \cite{Riddle74}. Using this kind model
in interpreting the origin of Type IIId bursts was suggested in
papers \cite{Abranin82,Abranin84}. However, the theoretically
calculated values were poorly coordinated with those UTR-2
measured. The assumption \cite{Abranin00} was therefore made that
the echo-component of stria-bursts is formed not merely due to the
reflection from a deeper layer of a spherically symmetric corona,
but also as a result of the refraction of radio waves on some
large-scale coronal structures. This hypothesis has been supported
in \cite{Afanasiev06} where the author has mathematically
simulated the shaping of Type IIId bursts. During the systematic
researches of  the  Sun with the UTR-2 array system and positional
observations with the two-dimensional heliograph for almost two
decades, more than 10 solar Type IIId burst storms were recorded.
It will be observed that Type IIId radio bursts are rather rare
events, unlike Type III bursts being most mass events of the
decameter sporadic solar radiation. With the complex observations
using the two-dimensional heliograph, in June 1984 during a
week-long Type III storm, about 1000 bursts were recorded at 25
MHz. The data obtained allowed the statistical sample manipulation
of decameter events by employing the cluster analysis
\cite{Stepanova95}. For the statistical procedure of sample
ordering and clustering, used were such parameters of bursts as
time delay between the two sequent bursts, burst du­ration at
half-power level, maximum intensity and degree of circular
polarization.

It should be recalled that way back in the mid-nineties, the
radioastronomy observations with the UTR-2 array were made only
near to several fixed frequencies: 10.0, 12.5, 14.7, 16.7, 20.0
and 24.8 MHz, while the broadbandness of this antenna system is
much greater. On the other hand, in the capacity of a recorder,
e.g. for the heliograph, the FTAP-2 facsimile set was used
recording the signal on a electrochemical paper with the dynamic
range making only 10-16 dB. Such a situation urgently demanded
extensive design improvement of the whole UTR-2 hardware facility.
When the UTR-2 antenna amplifier system had been upgraded in the
late nineties, the measurements became possible in the continuous
frequency band of 10-30 MHz and more \cite{Konovalenko97}.
Thereupon, an exigency of installation of appropriate recording
equipment arose. With that end in view, for multi-frequency
observations a 60-channel spectrum analyzer was put into service
which included 60 separate frequency-tunable within 10-30 MHz
radio receivers with the 4-10 kHz bandwidth. The next step was
applications of digital multichannel (1000 and more channels)
spectrum analyzers (digital receivers). These latter already
allowed covering the entire UTR-2 array frequency range with the
frequency separation of about 4 kHz, as well as recording on
digital storage media and perform varied signal computer
processing \cite{Kleewein97,Lecacheux98,Lecacheux2004}.

The very first configuration of the UTR-2 array based heliograph
was as follows. By means of a special beam control system of the
UTR-2 array it became possible to quickly scan the sky area under
investigation. Its image was recorded through signal injection
from the outputs of N-S and E-W (``East-West'') antennas into the
sum-difference device. The sum and difference signals were
amplified by individual receivers with square-law detectors. The
outputs of square-law detectors were in opposition, thus providing
the difference of their output voltages. This difference voltage
then fed the final recorder input. Actually, the radio astronomy
research efforts are basically carried out using the digital
receiver-recorder, being a digital spectropolarimeter (DSP)
capable of signal receiving and processing, and data recording on
the PC hard disk. Surprisingly, we still face the fact that though
the UTR-2 based heliograph was designed long ago and used in
diverse radioastronomy researches, its complete and thorough
description had been postponed because of ever-lasting upgrading
its hardware environment. For this long period, the heliograph has
been essentially reengineered, new possibilities for its
application appeared, and things have reached an exigent necessity
point to describe the heliograph development at long last review.

\section{Antenna System}\label{par3}
Using the available antenna system for the heliograph appliance
has in many respects determined its circuit design. The UTR-2
radio telescope operating within 10-30 MHz has been described in
detail in papers \cite{braude78,Men78}. Here in this paper, we
will only touch some of its array antenna features being
all-important of understanding the operation of the heliograph as
a whole.

\begin{figure*}
\centering
\resizebox{1.\columnwidth}{!}{%
  \includegraphics{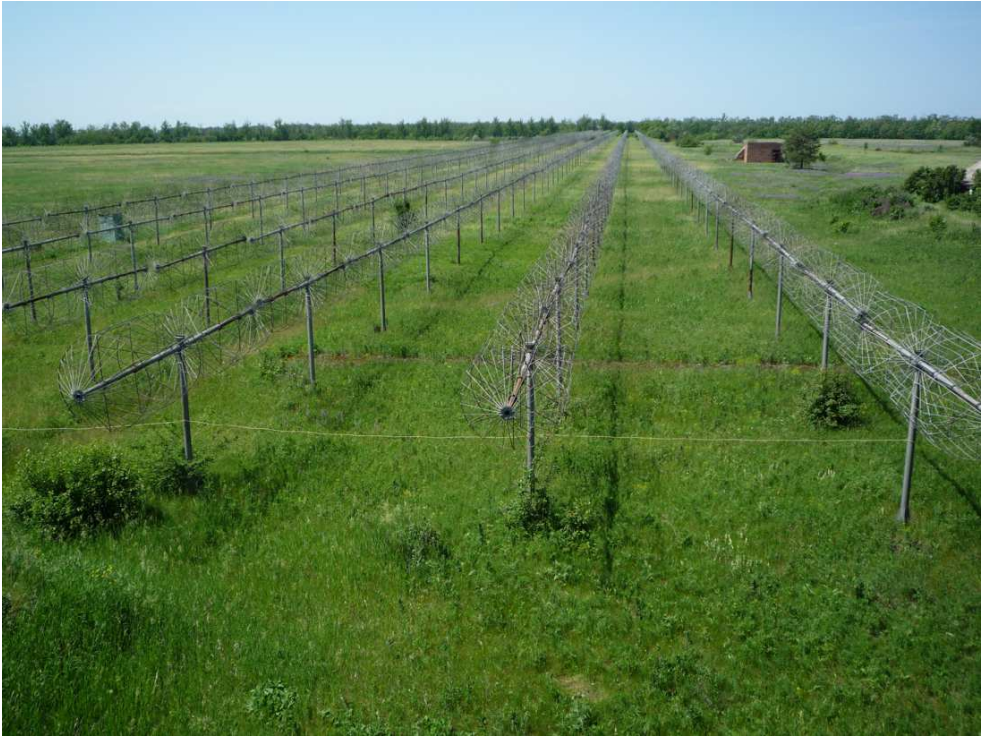}
}  \caption{East-West arm of the UTR-2 radio telescope.}
\label{fig}
\end{figure*}

The UTR-2 array antenna system consists of three rectangular array
arms: northern, southern and western. Each of the first two array
arms (N and S) includes 120 rows (each possessing 6 fat, therefore
broadband, horizontal dipoles) lined along the parallel. The E-W
array has 6 rows (each possessing 100 fat horizontal dipoles)
lined along the parallel (see Fig.~\ref{fig}). The horizontal
dipoles of all arrays are oriented along the east-west line. Along
the parallel, they are spaced by 9 m, and 7.5 m along the
meridian.

In the arrays with staged (otherwise stepped or storied) layout,
the signals of horizontal dipoles are summed and phased with
discrete cable delay lines. The signals are phased independently
by the two systems in two orthogonal coordinates, $V$ and $U$,
which are connected to equatorial coordinates with the relations
\begin{equation}
U = \cos\delta\,\sin t\,,\quad V = \sin\beta\cos\delta\cos t\
-\cos\beta\sin\delta\,, \label{eq1}
\end{equation}
where $t$ and $\delta$ are an hour angle and declination,
respectively, and $\beta$ is the geographic latitude of antenna
location.

The E-W antenna has one-stage phasing in $V$-coordinate and
three-stage phasing in $U$-coordinate, while the N-S antenna
one-stage phasing in $U$-coordinate and four-stage phasing in
$V$-coordinate. The circuit design allows signals' summation and
phasing in a manner that each array before the last phasing stage
appears to consist of four sections. But for all that, each
section has no common circuit elements with the others and
combines 1/4 of horizontal dipoles of the corresponding array. To
provide the best received signal-to-noise ratio, the signals from
all sections are amplified by array-distributed individual
amplifiers.

The amplified signals of eight sections of the N and S arrays are
phased, amplified by a line of antenna amplifiers, then each of
the signals furcating with the hybrid splitters into 5 channels.
The N-S antenna 5 beams, distributed in the $V$-coordinate at
$\Delta V = 1/150$, are shaped by phasing with the constant delay
of each of the channels of all eight sections and further
summation. In this case, the third beam is summed in phase, thus
shaping the central beam toward phasing $(U,V_3)$. Beam directions
along the $V$-coordinate are determined through the relation
\begin{equation}
 V_n = V_3 - \frac{n-3}{150}\,,\label{eq2}
\end{equation}
where $n$ is the number of a beam counted from the southward
direction. The directions of all 5 beams lie in one plane $V$, for
which $U_0 = {\rm const}$ and their hour angles, following the
relation (\ref{eq1}), are equal
\begin{equation}
 t_n = \arcsin(U/\cos\delta_n)\label{eq3}\,.
\end{equation}
Since in beam phasing, the time delays used are constant, the
declination distance between beams will be
\begin{equation}
 \Delta\delta \approx \Delta V/(\cos\beta\cos\delta + \sin\beta\sin\delta\cos
 t)\label{eq4}\,,
\end{equation}
where $t$ and $\delta$ are an hour angle and declination of
antenna phasing direction (in the zenith direction $\beta =
\delta, t = 0$ and $\Delta\delta = 23'$).

For shaping pencil beams, each E-W antenna output is multiplied by
one of the N-S antenna five beams. In this case, at 25 MHz, the
zenithal beams have sizes $28'\times 27\,' (t\times \delta)$ at
half-power width. As to the receiving of side-lobe radiation, it
is known to appear larger within the angles of factors of the
narrow sides of arrays in the mode of multiplication of the
T-shaped aperture arms. Its maximum power at the equi-amplitude
distribution of antenna currents makes about $22\%$, i.e. the same
level as that for the field array pattern. The side-lobe received
radiation contribution is controlled by varying the amplitudes of
the N-S antenna section currents.

The antenna has several types of phase shifters -- those with
different number of inputs and increments, as well as those with
summation of signals after and without phasing. The signal from
each input of the phase shifter comes upon a series of switched
binary-incremental cable time-delay lines. In essence, the series
are the separate phase shifters structurally making one shifter.
As switching elements, the high-frequency electromagnetic relays
are used with the pickup time of $\sim 20$ ms and cyclic life of
$\sim 10^5$ switchings. The phase shifters are remotely controlled
from the control panel either manually or using the PC
user-defined program.

The telescope coverage area is limited to $-1\leq V_0\leq 1$ and
$-7.5/9\leq U_0\leq 7.5/9$. In these limits, the direction­al
pattern of all stages of the both independent systems of array
phasing can be varied. The antenna employs asynchronously steered
patterns of array stages, i.e. their directions may differ. In
this case, the number of increments (and thus their values) of
phase shifters of array stages is selected reasoning from the
allowable decrease of directivity of antennas shaping the
corresponding directional patterns. Since with approaching the
array output the beam width of horizontal dipoles phased by one
phase shifter of a stage decreases, the number of increments in
the stage increases. The maximum number of increments has only
phase shifters of the last stages where the sections of E-W and
N-S antennas are phased. The E-W antenna has the phase shifter for
4 inputs and 10 bits, i.e. possessing the $2^{10} = 1024$ number
of increments over $U$; the N-S antenna having the phase shifter
for 8 inputs and 11 bits providing $2^{11} = 2048$ beam positions
over $V$ within the whole coverage sector.

The beams over $U$ and $V$ are convenient to number as $0\leq
\vert N_U \vert\leq 511$ and $0\leq \vert N_V \vert\leq 1023$,
attaching to them a negative sign if the beam is directed
accordingly eastwards and northwards (similarly to the $U$ and $V$
coordinates). Making allowance for the sign or quadrant choosing
is performed by the high-order digits of phase shifters, while the
rest 10 bits over $V$ and 9 bits over $U$ are intended for
recording of the absolute values of beam numbers. The beam numbers
over $U$ and $V$ are related to the phasing direction $(U,V)$ as
\begin{equation}
N_U = 512\frac{7.5}{9}U - 0.5\,\,{\rm sign}(U), N_V = 1024V -
0.5\,\,{\rm sign}(V)\label{eq5}\,.
\end{equation}
Here, the function ${\rm sign}(x)=\cases{\,\,\,\,1,\, {\rm if}\,\,
 x>0;\cr{-1},\,{\rm if}\,\, x<0.\cr}$

To optimize the magnitude of phase errors due to asynchronous beam
control, the phase shifters are de­signed so that the array beam
cannot be directed precisely to zenith. In the initial position,
the switching delays are switched off in all phase shifters.
However, due to the use in phase shifters of permanently engaged
time delays, the directions of patterns of array stages appear to
be displaced by the 1/2 of the corresponding stage increment.

Since in the general case the array beam direction is determined by
the product of directional pat­terns of stages possessing
noncoincident directions, the antenna beam phasing direction does
not, strictly speaking, coincide with either of UTR-2 antenna
factors. However, with taking into account that in both
coordinates the antenna beamwidth is more than 8-fold narrower
than the section directivity diagram, the phasing direction will
most closely coincide with the direction of the sharpest factor.
As the computing shows, the error in this case will not exceed
$0.6\,'$ over $t$ and $0.4\,'$ over $\delta$, which can be
neglected.

In that case, within the coverage sector, the radio telescope beam
can be positioned to any point of the sky accurate to no worse
than 1/2 of the magnitude of the sharp UTR-2 antenna factor
increment. In the equatorial coordinates, the error in beam
pointing, due to the control increment, will depend upon beam
direction. Within the angle sector where the solar observations
are made, it reaches a well noticeable magnitude (approximately
$3.5\,'$ and $2.5\,'$ over $t$ and $\delta$, respectively). The
account for a phasing increment of the antenna sharp factor is
therefore desirable.

To determine the true direction of the radio telescope (third)
beam phasing $(U_0,V_3)$, it is sufficient -- with the given
calculated phasing direction $(U,V)$ and using formula (\ref{eq5})
-- to find the corresponding magnitudes of $N_U$ and $N_V$. We may
round off them to the integral value using the formula
\begin{equation}
 N^* = {\rm int} ({\rm abs}(N))\cdot {\rm sign}(N)\,,\label{eq6}
\end{equation}
where ${\rm abs}(x)$ is the absolute value of $x$, and ${\rm
int}(x)$ is its integer part. Just these numbers of beams $(N^*_U,
N^*_V)$ correspond to the true phasing direction which is found
using the relations from (\ref{eq5}), namely
\begin{displaymath}
U_0 = ({\rm abs}(N^*_U)+ 0.5)/614.4 \cdot {\rm sign}(N^*_U + 0.5),
\end{displaymath}
\begin{equation}
V_3 = ({\rm abs}(N^*_V)+ 0.5)/1024 \cdot {\rm sign}(N^*_V +
0.5)\,.\label{eq7}
\end{equation}
In case of necessity, the direction of beams $(U_0,V_n)$ in the
equatorial coordinates can easily be found with the aforementioned
relations.

\section{Scanning System}\label{par4}
The UTR-2 scan sector allows one to observe the Sun for about $\pm
4^h$ around local noontime. Practically, however, due to
ionospheric screening at large zenith observation angles, growing
number of interfering broadcasting radiostations, and the
heliograph space resolution decrease, the scan sector is limited
over declination, $\delta_\odot \geq 10^\circ$, and over hour
angle, $\vert t_\odot \vert\leq 30^\circ$.

The five-beam UTR-2 operation and beam control in the two, $U$ and
$V$ coordinates, allows in principle to use this array without any
reconstruction as a 2D heliograph with either serial or
parallel-serial scan. Moreover, by having formed several beams of
the E-W arm in the $U$ coordinate, a parallel scan heliograph can
be built in which the maximum sensitivity is possible. However, no
advantage is taken of the opportunity so far seeing that a
sufficiently great amount of newly fabricated equipment is needed
in this case.

To observe both disturbed and quiet solar radio emission (for
which purpose the standard recording equipment of the radio
telescope is used) the parallel-serial scan system can be applied.
In this case, the output signals of all five beams of the N-S and
E-W antennas are simultaneously recorded by five parallel receiver
channels. As for the solar radio emission observations with the
heliograph employing the series principle, only one
receiver-recorder set is required as distinct from the previous
case.

In the both cases, the five-beam set located in the $V$ plane
sequentially holds a number of positions over $U$. It is achieved
by the E-W array pattern scanning. Since the beams are of about
the same size both in the $U$ and $V$ coordinates, obviously it is
expedient to take equal distances between them in these
directions, i. e. $\Delta U = \Delta V = 1/150$.

The gain in outlay for recording equipment -- in the case of using
the sequential as against the parallel-serial scan -- is
particularly appreciable in view of decisive expediency of
heliograph operation simultaneously at different frequencies.
Certainly, with lower frequencies, the observation efficiency
worsens because of the low space resolution. Even at 25 MHz, this
latter appears insufficient for the rather important measurements
of brightness distribution of solar decameter radiation sources.
However we must bear with this keeping in mind that in this case
we are dealing with the decameter wavelength range. Realization of
antenna systems for this range, whose sizes would be much greater
than these of the UTR-2 array, is rather problematic in itself.
Besides, the antenna size restriction takes place at decameter
wavelengths due to the ionosphere induced decorrelation of the
received signal.

\section{Heliograph Functional Scheme}\label{par5}
\begin{figure*}
\centering
\resizebox{1.\columnwidth}{!}{%
  \includegraphics{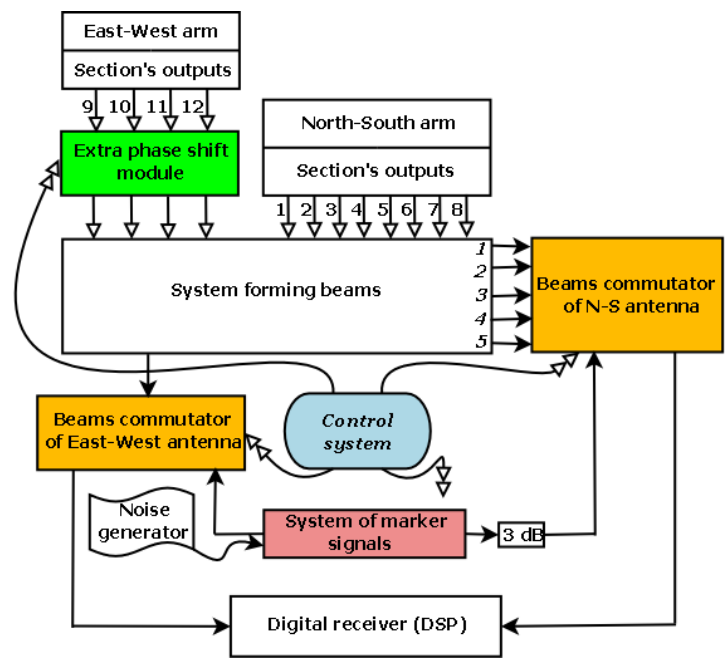}
}  \caption{The heliograph block diagram sketch.} \label{fig0}
\end{figure*}

With the above mentioned requirements taken into consideration,
the heliograph functional scheme was suggested and is shown in
Fig.~\ref{fig0}. The heliograph system incorporates the E-W and
N-S antenna arrays, too. An extra phase shifter (for fast
scanning) is sequentially connected to the E-W antenna section
outputs. The beam shaper phases and sums up the antenna section
signals, and then, using the splitters, it shapes a five-beam
directional pattern. The N-S and E-W beam arm switches connect in
turn the signals of a five-beam set, and also the noise signal
from the marker shaper to the DSP. The marker shaper is integrated
into the noise generator and generates the right ascension
separating markers as well as markers of end (beginning) of image
on heliograms. The signal strength of the marker of end
(beginning) of image is higher than the separating marker between
the five-beams due to the incorporated 3 dB attenuator. This
permits avoiding difficulties in analyzing the heliograms. The
control unit ensures the beam switching, markers shaping and phase
shifting of fast scanning.

In what follows, we will dwell on some important features of the
considered device.

\subsection{Heliogram Repetition Rate}

The heliograph image repetition rate should be rather high so that
the burst evolution could be observable. However, a too high
repetition rate will reduce the receiver sensitivity and
complicate the recorder design. The maximum image repetition rate
was taken 4 fps. At such a repetition rate, the heliograph allows
observation of the bursts with duration of $\geq$ 1 s that is
quite sufficient for the given range of radio observations. In
this case, the receiver sensitivity remains high enough, while the
recorder simple and compact. That high frequency of a heliograph
image repetition rate is not always required. It can be lower as
well (e.g. for observations of the quiet Sun radio emission). Such
a possibility is provided by the change of clock rate in the
control unit.

\subsection{Polarization and Frequency Passband}
While observing the sporadic radio emission of the Sun, in
addition to the bursts' position measurement, determining their
polarization is of great interest, too. However because of the
fact that the UTR-2 array antennas are designed to receive only
the linear polarization, the polarization measurements are
impossible.

For the better signal-to-noise ratio, the frequency passband
should be wide enough. The upper limit of the passband width is
determined by the frequency band of the observed phenomenon
($\sim$ 1 MHz). On the other hand, the noise density and intensity
from the short-wave broadcasting stations are extremely great at
the decameter wavelengths. For this reason, the passband width
cannot be greater than several kilocycles. Therefore the entire
recording band of 10-30 MHz can be conveniently subdivided by the
channels of $\sim$ 4 kHz. As will be shown below, the DSP
perfectly copes with such a task.

\subsection{Switches}
In a heliograph with sequential shaping and recording of image
elements, the receiver-recorder is sequentially connected to the
outputs of the N-S antenna five beams. In the end of each beam
scanning cycle, instead of antennas, the auxiliary noise generator
-- which helps forming markers on the heliogram -- is connected to
the receiver-recorder.

The heliograph incorporates two switches: the N-S antenna beam
switch and the E-W antenna switch. The former is used for the
sequential connection of the out­puts of the N-S antenna five
beams and the auxiliary noise generator to one of the
receiver-recorder inputs. The E-W antenna switch connects to the
second input of the receiver-recorder either the E-W antenna
output, or the auxiliary noise generator. The E-W antenna remains
connected to the receiver-recorder for the whole beam scanning
period. In the end of each scanning cycle of the kind, the
auxiliary noise generator is connected, instead of antennas, to
the both inputs of the receiver-recorder.

Thus, the E-W antenna switch has two inputs and one output (the
first input is fed by the E-W antenna output signal, while the
second by the signal for forming a marker of the end of scanning
five beams). The N-S antenna beams switch has six inputs and one
output (five inputs for connection of outputs of N-S antenna five
beams, while the sixth input is fed by the signal forming a marker
of the end of scanning beams and transfer to the next scanning
cycle). While the E-W antenna pattern rests in the predetermined
position, several cycles of scanning N-S beams can be performed.
The signal from the first input of the E-W antenna switch feeds
the switch output for the time of scanning five-beam set of the
N-S antenna by its switch. The second input signal feeds the
output when the sixth input beam switch is switched on. Thereby,
in this case the signals from the N-S and E-W antennas outputs are
disconnected from the inputs of the receiver-recorder, and instead
the signals for forming a marker, which indicates the end of
scanning five beams, are connected.

Concurrently with the eighth marker of end of scanning five-beam
set, the end-of-image marker should be shaped to avoid possible
errors in determining the frame boundary in time. The end-of-image
marker can be combined with the time marker provided the UTR-2
array's code-switch time is matched with the E-W array pattern
switching time.

The E-W antenna arm switch consists of two, while that of the N-S
arms of six identical cells in which the crystal fast-response and
long-lifetime diodes are used as switching elements. For removal
of reflections from the unloaded inputs of disconnected channels,
all of them are loaded by the matched loads of 75$\Omega$ for a
disconnection period. The attenuation in open channels of switches
makes $\sim$ 1.2-1.5 dB and is largely determined by losses in
switching diodes. Structurally, the N-S and E-W antenna beam
switches are made as the two separate printed-circuit boards. Each
switch has a monitoring circuit that allows fast detection of
malfunctions to appear in operation.

\section{Fast Scan Phase Shifter Design}\label{par6}
The heliograph beam control system has two functions: a rather
slow Sun tracking and fast scanning over the heliograph scan area.

Though the UTR-2 radio telescope was designed as a
multi-functional instrument, its phasing system does not allow
long observations with fast beam scanning because of
electromagnetic relays used in it. That is why the heliograph
utilizes only the UTR-2 phasing system for tracking the Sun with
the beam switching rate of several minutes. As for the fast
scanning, this function is executed by a specially designed
phasing system.

This latter, as has already been noted, should discretely  change
the E-W array pattern position over $U$. The complexity of the
beam fast scan system is determined by the point of its connection
to the UTR-2 antenna phasing circuit seeing that it should embrace
all the subsequent stages and never do any of the previous ones.
From the point of simplicity, it is expedient to be connected as
near as possible to the array output.

On the other hand, the point of its connection is determined by
the heliograph scan sector in the $U$ coordinate. Its width should
be comparable to (or less than) the one of the antenna beam
patterns of the elements which are phased by this stage. On the
last stage, where the E-W arm section output signals are phased,
the field beamwidth makes about $\sim 3^\circ$  at 25 MHz. Such a
scan heliograph sector over $U$ appears to be quite sufficient
even for observation at 12 MHz, where the solar corona radio
diameter makes about 1.5$^\circ$.  For switching on the fast scan
system at the last phasing stage, only one 4-input phase shifter
is required.

The beam fast scan phase shifter is circuit-connected between the
section outputs and corresponding inputs of the last E-W arm
phasing stage. In this case, the section beam pattern appears to
be a scanning beam envelope.

Structurally, the fast scan phase shifter is similar to that used
in the UTR-2 phasing system \cite{bruk78}. The difference lies
only in that its coaxial cable lengths are switched with the diode
keys which have larger service life and operating speed ($\sim$
0.5 ms). The phase shifter integrates four lines, each being a
three-digit digital-binary time delay line. For short, we
hereinafter will use the abbreviation of this phase shifter as the
type FV4-3 ($\Phi$B4-3 in Russian transcription). Among themselves
these strips differ only by the values of switched time delays.
The control circuit for the isobits of all strips is connected in
parallel.

Using the diode keys and the coaxial cable lengths as the time
delays ensures broadbandness for such a phase shifter. The signal
phase deviation from the calculated value does not exceed
5$^\circ$ at 25 MHz. The initial cable lengths in the phase
shifter strips ($\ell_0$) are selected so that in the starting
position (when all delays are switched off) the E-W arm beam
appears to be located eastwards to the $U_0$ phasing direction. As
the phase shifter passes through all its 8 positions, the beam
with a step of $\Delta U = 1/150$ moves westwards.

It is natural to take the scan sector being located symmetrically
with respect to the $U_0$ phasing direction. As the number of
phase shifter positions is even, neither of its positions does
coincide with the $U_0$. The five-beam set scanning over $V$
sequentially takes the values over $U$, so that
\begin{equation}
 U_m = U_0 + \frac{m-4.5}{150}\,,\label{eq8}
\end{equation}
where the beam position number $1\leq m\leq8$. In this case, no
N-S antenna phasing over $U$ is required as its pattern width is
greater versus the E-W antenna scan sector. Only the cable length
to compensate the average electrical length of the fast scan phase
shifter is additionally connected to the N-S antenna.

To calibrate the heliograph receiver-recorder chain, the
electrical lengths of the E-W and N-S antennas should be
equalized. As for $U_0 = V_0$ and for any $m$, according to
expression (\ref{eq8}) we have $U_m\not=0$, the additional cable
delays should be used for compensation. In operation mode they are
switched off.

\section{Calculation of Cable Lengths for FV4-3}\label{par7}
\begin{figure*}
\centering
\resizebox{0.8\columnwidth}{!}{%
  \includegraphics{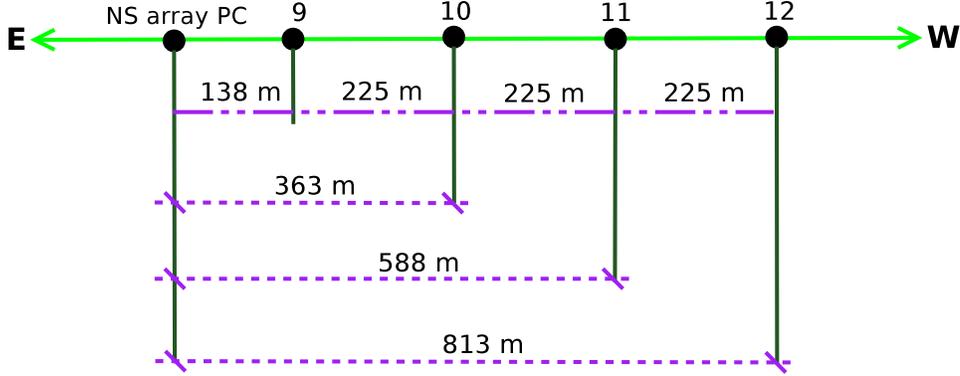}
}  \caption{Spatial position of the N-S array phase centers (PC)
and E-W arm sections (figures 9, 10, 11, and 12 denote section
numbers).} \label{fig1}
\end{figure*}

In heliograph operation, the maximum additional UTR-2 beam
deflection over $U$, with respect to the direction given by the
array phasing system, makes $\Delta U = \pm 7/300$. The phase
center of the heliograph fast scan phase shifter FV4-3 is combined
with the N-S array phase center (being a phase shifter with the
displaced phase center). Fig.~\ref{fig1} shows the spatial
position of N-S array phase centers and E-W arm sections.

\begin{figure*}
\centering
\resizebox{0.8\columnwidth}{!}{%
  \includegraphics{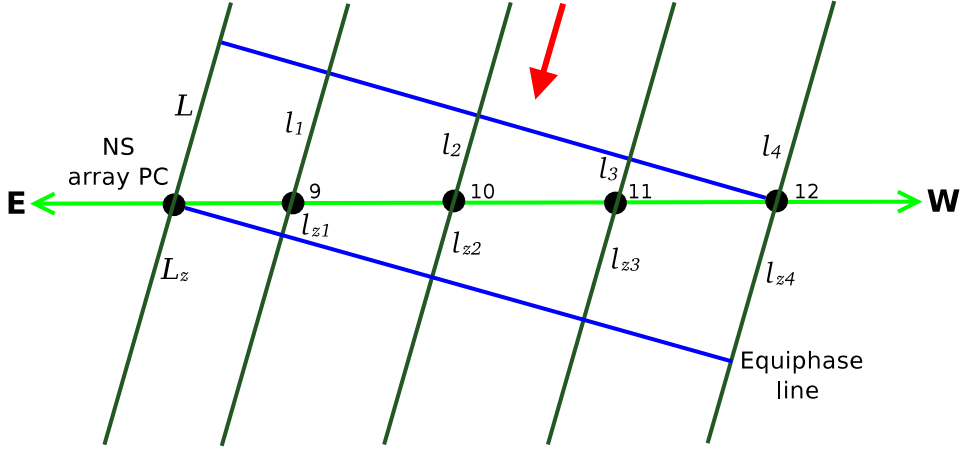}
}  \caption{Compensation of path differences by connecting some
additional delay-line lengths ($L_z$, $l_{z1}$, $l_{z2}$,
$l_{z3}$, $l_{z4}$).} \label{fig3}
\end{figure*}

To equalize the phases of received signals at the outputs of all
antenna system elements, the natural path differences $L$, $l_1$,
$l_2$, $l_3$, and $l_4$  are compensated with additional delay
line lengths (cables) connected in the FV4-3 (see
Fig.~\ref{fig3}). The electrical lengths of additional delay lines
have the values $L_z$, $l_{z1}$, $l_{z2}$, $l_{z3}$, $l_{z4}$, and
in this case $L_z=l_4$, $l_{z1}=L-l_1$, $l_{z2}=L-l_2$,
$l_{z3}=L-l_3$, $l_{z4}=L$.

\begin{figure*}
\centering
\resizebox{0.8\columnwidth}{!}{%
  \includegraphics{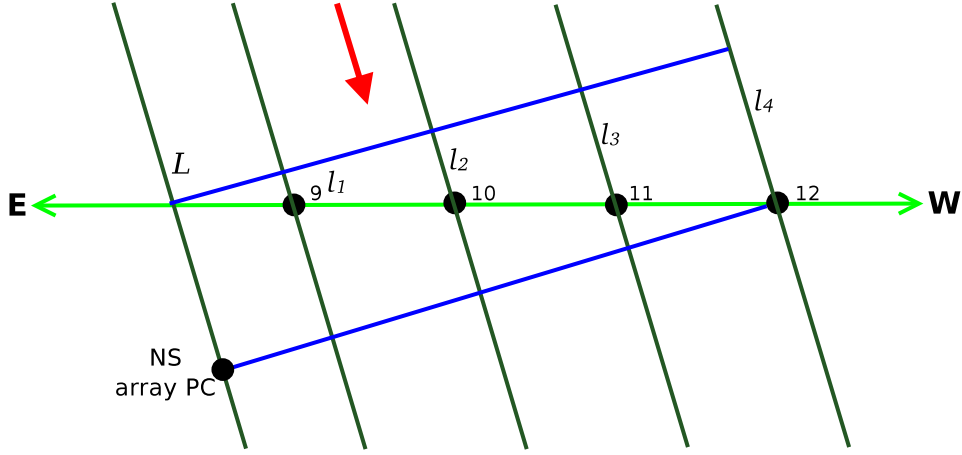}
}  \caption{N-S array phase center (PC) position displacement by
connecting some permanent length $L_z$ for the beam deflection
eastwards.} \label{fig4}
\end{figure*}

For the beam eastward deflection and for the unchanged phase
center position in the N-S array center, the path differences
should be compensated by connection of negative lengths in the E-W
array, which is impossible. Therefore the phase center is
displaced by connection of the additional length $L_z = l_4$
(Fig.~\ref{fig4}). This additional length remains permanently
connected with the beam deflection both eastwards and westwards,
while the compensating lengths in the FV4-3 change respectively.

As the maximum heliograph beam deflection over $U$ is $\Delta
U_{max} = \cos\alpha = 7/300$ and the maximum distance between the
extreme phased elements $L_m = 138 + 3 \cdot 225 = 813$ m, then
$L_z = L_m\cdot\cos\alpha = 813\cdot3/700 = 18.97$ m. For the
eastward deflection, the phasing electrical lengths are as follows
\begin{eqnarray}
L_0&=&L_z\,,\nonumber\\
l_{01}&=&225\cdot3\cdot7/300 = 15.75 {\ \rm m}\,,\nonumber\\
l_{02}&=&225\cdot2\cdot7/300 = 10.5 {\ \rm m}\,,\nonumber\\
l_{03}&=&225\cdot7/300 = 5.25 {\ \rm m}\,,\nonumber\\
l_{04}&=&0.\nonumber
\end{eqnarray}
For the westward deflection, the phasing electrical lengths make
\begin{eqnarray}
L_{z}&=&0 + l_{z4} = 0+813\cdot7/300 = 18.97 {\ \rm m}\,,\nonumber\\
l_{z1}&=&138\cdot7/300 + L_z = 3.22 + 18.97 = 22.19 {\ \rm m}\,,\nonumber\\
l_{z2}&=&(138 + 225)\cdot7/300 + L_z = 8.47 + 18.97 = 27.44 {\ \rm m}\,,\nonumber\\
l_{z3}&=&(138 + 2\cdot225)\cdot7/300 + L_z = 13.72 + 18.97 =
32.69 {\ \rm m}\,,\nonumber\\
l_{z4}&=&(138 + 3\cdot225)\cdot7/300 + L_z = 18.97 + 18.97 = 37.94
{\ \rm m}\,.\nonumber
\end{eqnarray}
Then, the lengths which are switched in the FV4-3 operation should
respectively have the values
\begin{eqnarray}
L_k&=&L_z - L_0 = L_z - L_z = 0\,,\nonumber\\
l_{k1}&=&l_{z1} - l_{01} = 22.19 - 15.75 = 6.44 {\ \rm m}\,,\nonumber\\
l_{k2}&=&l_{z2} - l_{02} = 27.44 - 10.5 = 16.94 {\ \rm m}\,,\nonumber\\
l_{k3}&=&l_{z3} - l_{03} = 32.69 - 5.25 = 27.44 {\ \rm m}\,,\nonumber\\
l_{k4}&=&l_{z4} - l_{04} = 37.94 - 0 = 37.94 {\ \rm m}\,.\nonumber
\end{eqnarray}
\begin{figure*}
\centering
\resizebox{0.8\columnwidth}{!}{%
  \includegraphics{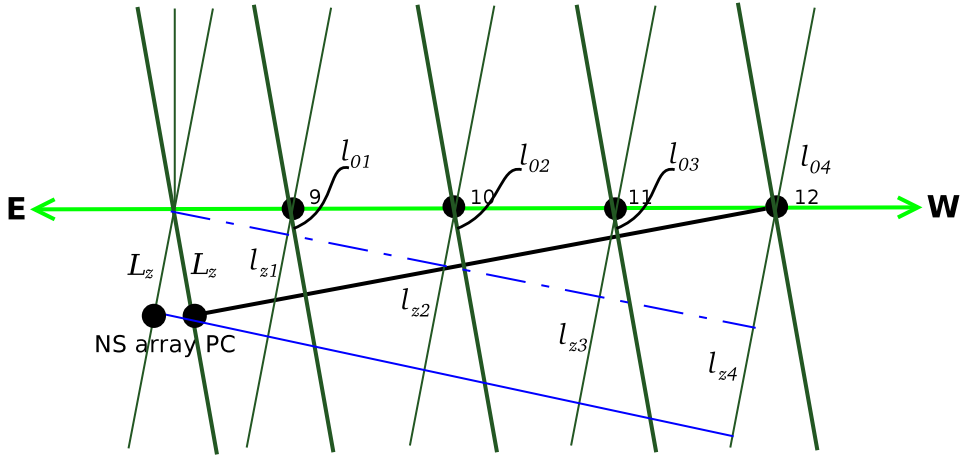}
}  \caption{Schematic representation of the UTR-2 array pattern
position for the extreme western and eastern positions: $L_{z}$,
$l_{01}$, $l_{02}$, $l_{03}$, and $l_{04}$ are phasing lengths for
the beam deflection toward east; $L_{z}$, $l_{z1}$, $l_{z2}$,
$l_{z3}$, and $l_{z4}$ are phasing lengths for the beam deflection
toward west with allowance for a permanently connected length
$L_{z}$.} \label{fig5}
\end{figure*}
The delay lines are made by employing the binary principle, each
having three switched bits. The delay values in the bits are
interrelated as 1 : 2 : 4. Besides the switched elements, the
delay lines include also the non-switchable elements  $l_{01}$,
$l_{02}$, $l_{03}$ and $l_{04}$. Such a structure of delay lines
allows to discretely set the heliograph beam into eight positions
in the $U$ coordinate with a step 1/150 within $-7/300$ to
$+7/300$ (Fig.~\ref{fig5}). All line bits are switched
synchronously.

With the selected structure of delay lines, the length of the
least-significant bit $l_{\rm I} = l_k/7$  while that of the
second bit $l_{\rm II} = 2\, l_k/7$, and that of the third bit
$l_{\rm III} = 4\,l_k/7$. The permanent length connected in the
N-S array is equal to $L_0 = 18.97$ m. All the FV4-3 elements are
made of the RK 75-9-12 (PK 75-9-12 in Russian transcription) type
coaxial cable having the characteristic impedance of 75 $\Omega$
and wave shortening coefficient of $k$ = 1,51. Thereto, for the
FV4-3 connection into the UTR-2 phasing system, each delay line
should additionally possess two special cables with total length
of $l_t = 5$  m. The total length of special cables in each line
is added to the length of a non-switchable cable length $l_0$. The
Table~\ref{delay2} shows the corrected FV4-3 cable lengths with
allowance for the wave shortening coefficient and some processing
aids into non-switchable delay lines. The special cable length is
added also to the N-S antenna cable, thus $L_0 = 17.56$ m.

\begin{table}
\centering \caption[]{Phase shifter cable delay line lengths (in
meters), $l_{\rm 0}$ is the initial length, $l_{\rm I}, l_{\rm
II}$, and $l_{\rm III}$ are switched lengths.\\ } \label{delay2}
\begin{tabular}{c|cccc}
\hline & \multicolumn{4}{c}{E-W array arm section}\\\cline{2-5}
\raisebox{1.5ex}[0.3cm][0.3cm]{Length, m} &
\multicolumn{1}{c}{$\sharp 9$} & \multicolumn{1}{c}{$\sharp 10$} &
\multicolumn{1}{c}{$\sharp 11$} & \multicolumn{1}{c}{$\sharp 12$}
\\ \hline $~~l_{\rm 0}$ & 10.43 & 6.95 & 3.48 & 0 \\ $~~l_{\rm I}$ & 0.61
& 1.60 & 2.60 & 3.59 \\ $~~l_{\rm II}$ & 1.22 &
3.20 & 5.19 & 7.18 \\ $~~l_{\rm III}$ & 2.44 & 6.41 & 10.38 & 14.36 \\
\hline
\end{tabular}
\end{table}

The phase shifter attenuation with the delay line length variation
from 0 to 38 meters changes within 3.3 to 4 dB and is mainly
caused by the losses in switching diodes.

\section{The Heliograph Scan Sector Format}\label{par8}
\begin{figure*}
\centering
\resizebox{1.\columnwidth}{!}{%
  \includegraphics{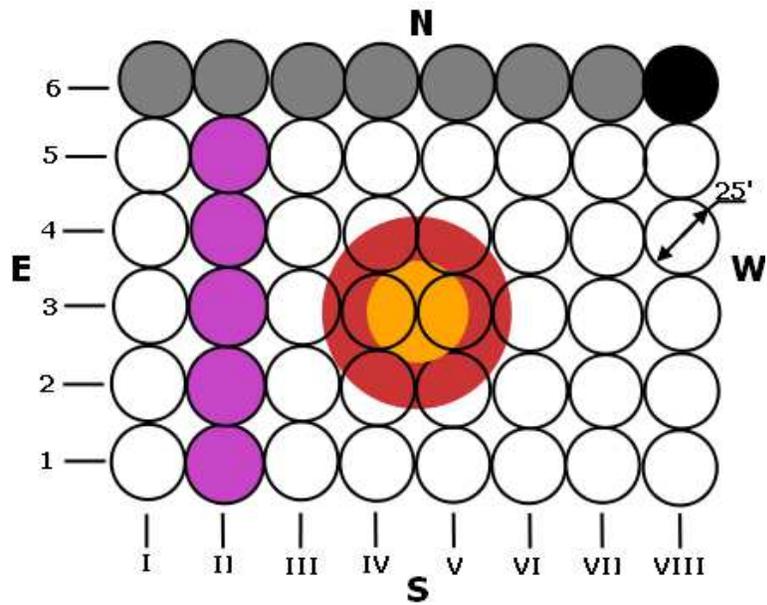}
}  \caption{A heliogram field  formed by the heliograph with the
resolution of  5 $\times$ 8 pixels. Angular size of each element
in the $t$ and $\delta$ coordinates makes 25$'$. Grey and black
circles are separating markers.} \label{fig8}
\end{figure*}

So, any heliogram is formed owing to the fact that with the fast
scan phase shifter, the five pencil beams spaced in the $V$
coordinate take sequential positions in the $U$ coordinate. Thus,
the full heliogram of the heliograph scan region is a rectangular
matrix of 5 rows and 8 columns (making totally 40 elements) spaced
over hour angle and declination by 25$\,'$ (see Fig.~\ref{fig8}).
The image angular size over hour angle makes $\sim 3.3^\circ$ at
25 MHz.

Such an element number ratio over declination and right ascension
allows for the fact that in the close-to-solar-minimum years the
equatorial diameter of the solar corona decameter radioimage is
almost 1.5-fold exceeding the polar one. Besides, the decameter
solar emission sources, as is known, are also crowded in the solar
equatorial plane. At the same time, a whole series of the problems
does not require that large scan sector over $U$. Therefore the
possibility for changing the number of elements in lines is
provided, thus allowing to increase the frame rate.

In the $U$ and $V$ coordinates, the image elements form a
rectangular frame. However the observed data are more preferable
to be represented in equatorial coordinates. And from the
viewpoint of their obviousness it would be better that the beam
antenna pointing to solar disk remained unchanged. Though, as was
already noted, in the mode of heliograph Sun's tracking the
distances between the beams change both over $t$ and $\delta$. To
avoid some extra distortions related to fast scanning over $U$,
the third beam declination should remain constant for all
five-beam set positions over $U$. This can be ensured at each
$U$-step by appropriate N-S array phasing over $V$. For each
$U_m$, the $V$ value should be such that the right-hand side of
expression
\begin{equation}
\sin\delta_{m3} = -V\cos\beta + \sin\beta\sqrt{1-U_m^2-V_3^2}
\label{eq9}
\end{equation}
should be constant. Here $\beta$ is the UTR-2 array geographic
latitude.

In order to cope with the above task, an 8-input three-bit phase
shifter, as well as more sophisticated phase shifter control
equipment, are required. Therefore the heliograph is scanning the
five-beam sets only by the E-W arm pattern steering over $U$.

The UTR-2 array solar tracking system is carried out by
superimposition of the scan area center with the solar disk
center. Note that the rectangular raster is preserved only for
near-meridian operation. In the other cases, the raster shape may
essentially differ from the rectangular one. This phenomenon is
caused by the fact that in antenna pattern steering, the $U$ or
$V$ coordinate changes at fixed values of the other coordinate,
while the increment of coordinates $\Delta U$ and  $\Delta V$ are
taken constant and independent on the $U$ or $V$ coordinate
values. This raster shape feature should be taken into
consideration in the observed data reduction, otherwise noticeable
errors may result in determination of coordinates of bursts in
heliograms.

In overwhelming majority of cases, the burst source center
position with respect to the imaging array  (Fig.~\ref{fig8}) is
such that the burst response occupies several image elements. Each
element of brightness is determined by its spacing from the burst
source center. In data processing, while determining the burst
coordinates, the image elements intensity ratios are measured and
compared with the imaging array. The imaging array coordinates are
calculated beforehand. Such a technique allows determination of
the burst angular coordinates to an accuracy of $\sim$ 5$\,'$.

\section{Control System}\label{par9}
The operation of fast scanning phase shifters, switches, marker
schemes and attenuators is accomplished by the control unit in
which the appropriate control pulse sequences are formed.

With the ``Reset'' command to the control circuit of the FV4-3
phase shifter (being a four-channel switch of binary-discrete
three-bit time delay cable lines which are sequentially connected
to the four outputs of the E-W array sections), the E-W array
pattern is hopping to the extreme ``eastern'' position (position
I, code 000). For this purpose, all the FV4-3 cable delays are
disconnected (operating in minimum-delay mode). During heliograph
operation, the E-W arm pattern position can discretely change
moving westwards (totally 8 positions). In the extreme ``western''
position (position VIII, code 111), the time delay cable lengths
are maximal.

The FV4-3 unit has four switchboards. Their circuits are
identical, only the lengths of switched cables are different. The
control pulses are sent simultaneously to all four switchboards on
the analogous contacts. To connect the required delay line
($\ell$, 2$\ell$ or 4$\ell$), the cutoff voltage drop should be
applied to the appropriate contacts thus locking the switching
diodes connected in parallel to the delay cable, and
simultaneously the zero voltage drop be applied to the conjugate
contacts to unlock the switching diodes connected in series to the
delay cable. For the delay cable disconnection, the switching
voltage should be applied to the appropriate control inputs in the
reverse order. In this case, the diodes connected in parallel to
the delay cables are on, while those connected in series to the
delay cable are off.

For the maximum frame repetition rate (to 4 fps) and number of
image elements (48 elements per frame), the clock rate which
controls operation of switches is equal to 200 Hz. A rather high
switching frequency has governed the choice for switching elements
of crystal diodes possessing high operating speed and large
service life.

To eliminate the reflections from non-loaded inputs of
disconnected channels, all of these are loaded with 75 $\Omega$
resistors. The attenuation in the open channel of the switch makes
$\sim$ 1 dB and is mainly determined by the diodes resistance
losses.

After the recent control unit design improvement, new modes for
the master clock pulses, viz. 0.2, 0.4 and 200 Hz, appeared.
Availability of several operational modes has allowed choosing
different forming time of the heliograph frame. Earlier, the
100/200 Hz frequency oscillator was synchronized by the electrical
grid frequency, inasmuch as the mechanical system of line scanning
of a recording device -- being a facsimile receiver -- was
synchronously driven. Owing to implementation of a new
receiver-recorder (DSP), the necessity in such an oscillator
ceased to have significance. A new oscillator variant is
crystal-controlled that has allowed to noticeably improve stable
operation as a whole. Besides, the sum-difference unit -- earlier
always having permitted to form a heliogram pattern pencil beam
using the N-S and E-W array signals -- became quite unnecessary.
At present, the DSP perfectly fulfills this function, but the
equipment will be considered more comprehensively in the next
section.

The beam switch and E-W array switch control pulses are formed
with the   frequency divider  and the decoder. The fast scanning
phase shifters control pulses are formed with the   frequency
divider.

The control unit also comprises the display circuit to which the
signals arrive from the monitoring circuits of phase shifters and
switches. The monitoring and display circuits allow ones to fast
determine the serviceability of these devices and quickly found
and remedy the breakages.

\section{DSP Application in Heliograph}\label{par10}
It will be observed that as far back as the late nineties, the
observations with the UTR-2 array were made only at six fixed
frequencies. Such a situation hindered from obtaining a more
comprehensive information on the recorded events, as well as
resulted in an appreciable loss of advantage against
multi-frequency observations. In recent years, however, the scope
for the radio telescope receiver-recorder has been essentially
expanded.

A standard approach in studying non-stationary random signals can
be the instantaneous spectral analysis \cite{bendat}, which using
allows one to transform the input signal into a two-dimensional
(time-and-frequency) spectrogram. The very first broadband digital
spectral receivers for the low-frequency radio astronomy employing
the real time Fourier-analysis (DSP, Robin) were developed under
the joint France-Austria project \cite{Kleewein97,Lecacheux98} and
successfully introduced at the NDA (France) and UTR-2 (Ukraine)
radio telescopes \cite{Lecacheux2004}. These latter have permitted
obtaining a great bulk of new results, including those in studies
of the Sun, which are partially presented in papers
\cite{melnik04,melnik05,chernov,konov07}. The further progress of
digital and computer facilities, information and telecommunication
technologies has allowed creation of spectral receivers with
substantially improved parameters \cite{Vinogradov07, Ryabov10}.
In these latter, the input signal is digitized with the high-speed
analog-to-digital converter (ADC), being then processed by the
FPGA matrix that results in determination of the signal spectra in
the time domains where the signal can be considered as
quasistationary. The spectra obtained are then written as a number
matrix (frequency channels -- timing of the corresponding spectra
origin) on the PC hard disk. The software developed provides for
the graphic user interface in order to choose the data recording
mode, data mapping after the respective real time transformations,
as well as monitoring a number of other spectrum analyzer
parameters.

\begin{table} \centering \caption[]{The DSPZ key features.}\label{dsp2}
\begin{tabular}{p{7cm}p{5cm}}
\hline
\footnotesize Sampling frequency(MHz) & \footnotesize 66 \\
\footnotesize Operating frequency range(MHz) & \footnotesize 8--32 \\
\footnotesize Total number of frequency channels & \footnotesize 8192\\
\footnotesize Frequency resolution(kHz) & \footnotesize 4 \\
\footnotesize Time resolution(s) & \footnotesize от 0.2$\cdot 10^{-4}$ -- 1 \\
\footnotesize Dynamic range (dB)& \footnotesize 117 \\
\hline
\end{tabular}
\end{table}

\begin{figure*}
\centering
\resizebox{1.\columnwidth}{!}{%
  \includegraphics[width=1.\textwidth]{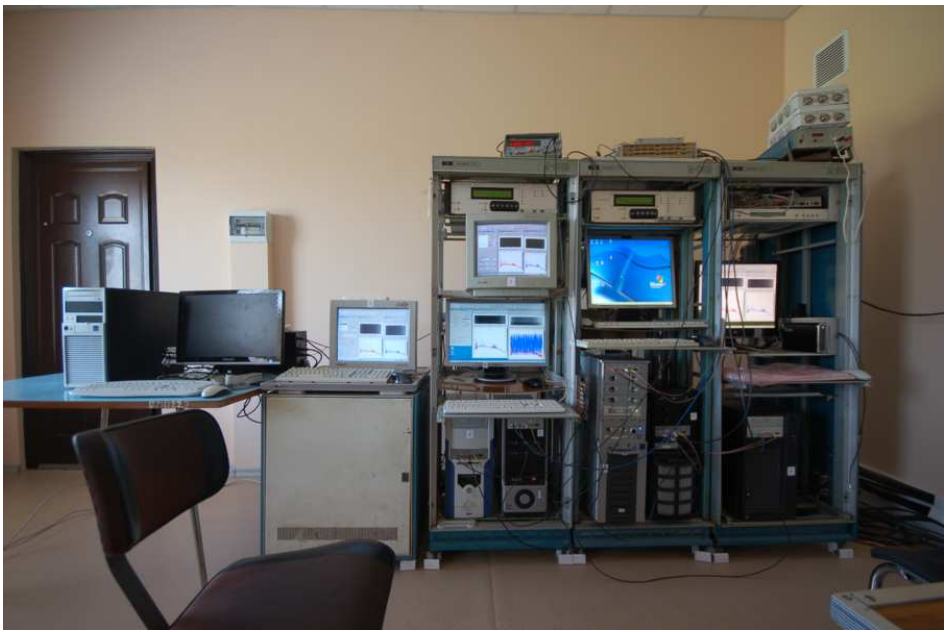}
} \caption{The UTR-2 DSPZ spectrum analyzers along with control
and data acquisition PCs are shown. } \label{fig8a}
\end{figure*}

At present, the DSPZ spectrum analyzer (see Fig.~\ref{fig8a}) is
used as a heliograph's receiver-recorder. Its key features are
presented in Table~\ref{dsp2}. It can operate in three different
modes, called ``Waveform'', ``Spectrum'', and ``Correlation''. In
the first one, the data flow from the output of one and/or two ADC
channels is recorded on the PC storage device without any
transformations or analysis. This is the simplest data recording
method whose subsequent processing is entirely performed by the
user. But in this case, an enormous data flow will be processed
(or recorded). This flow makes about 264 Mb/s because the data
digitization frequency is equal to 66 MHz. In the ``Spectrum''
mode, the window fast Fourier transform is applied. This procedure
translates the scalar input signal into a two-dimensional
spectrogram   otherwise called a dynamic spectrum. Function   is
an average spectral power of the window signal,
\begin{eqnarray}
 P_n(t,f)&=&|X_n(t,f)|^2,\nonumber\\
 X_n(t,f)&=&\frac{1}{T}\int\limits_{t-T/2}^{t+T/2}x_n(\tau)g(\tau-t)\exp(-i2\pi
 f\tau)d\tau\,,\label{eq10}
\end{eqnarray}
where $n$ = 1, 2 is the input channel number, and $g(\tau-t)$ is
the window function. The time window $T=2^m\Delta t$ is associated
with the interval between two sequential samples of the input
signal $\Delta t$ and the choice of number of spectral analysis
channels. The data flow rate makes about 327 kb/s with the time
resolution 100 ms. In this mode, the two input signals are treated
separately from each other, therefore at the output we have two
independent data flows, though the data lack the information about
the phase of signals. The ``Correlation'' mode allows broader
possibilities for the data analysis. Note that both in the
``Spectrum'' mode, as well as in the ``Correlation'' mode too, the
spectral density matrix can be determined in the form
\begin{displaymath}
{\bf Z} = \Bigg[\begin{array}{cc}
Z_{11} & Z_{12}\\
Z_{21} & Z_{22}
\end{array} \Bigg]\,
= \Bigg[\begin{array}{cc}
P_1 & Z_{12}\\
Z_{21} & P_2
\end{array} \Bigg]\,,
\end{displaymath}
where $P_1,P_2\in {\bf R}_+\cup 0$ and $Z_{12},Z_{21}\in {\bf C}$
mind though that it can be used completely only in the
``Correlation'' mode. This matrix is useful in the cross-spectrum
analysis, when through the complex spectra of signals
$x_{1,2}(\tau)$ their cross-spectral density is written as
\begin{displaymath}
K(t,f) = Z_{12} = \langle X_1(t,f)X_2^\ast (t,f)\rangle,
\end{displaymath}
where the magnitudes $X_{1,2}(t,f)$ are calculated according to
(\ref{eq10}), symbol ``$*$'' means complex conjugate value, while
the angular brackets denote the averaging over intervals $T_i$.
The cross-spectrum analysis allows one to test the signals
identity. Function $K(t,f)$ is complex and contains the
information about the phase relationship of received signals.
Therefore it is convenient to be presented exponentially as
\begin{displaymath}
K(t,f) = |K(t,f)|\exp[-j\Phi(t,f)],
\end{displaymath}
where the absolute magnitude and phase angle are determined by the
relations
\begin{equation}
|K(t,f)| = \sqrt{[{\rm Re}({\it K}(t,f))]^2 + [{\rm Im}({\it
K}(t,f))]^2},
\end{equation}
\begin{equation}
\Phi(t,f) = \arctan\left(\frac{\rm Im (\it K(t,f))}{\rm Re (\it
K(t,f))}\right),
\end{equation}
and $j$ is the imaginary unit. The real part of the cross-spectral
density function $\rm Re (\it K(t,f)) = C(t,f)$ is called a
cospectrum, and its imaginary part $\rm Im (\it K(t,f)) = Q(t,f)$
a quadrature spectrum. To come to the point of a cross-spectral
function, suffice it to mention that the cospectrum $C(t,f)$ makes
a contribution of oscillations of different frequencies to the
general cross-spectrum for the zero phase shift between two
signals. Meanwhile, the quadrature spectrum shows the contribution
of different harmonics into the general cross-spectrum, when the
harmonics of the first signal $x_1(\tau)$ are late for a quarter
of period with respect to the corresponding harmonics of the
second signal $x_2(\tau)$. In observations at the UTR-2 array, one
input of DSPZ is fed with the N-S antenna signal, while the other
one with the E-W one, that allows to eventually shape a
pencil-beam antenna pattern for the heliograph.

The ``Correlation'' mode possesses yet not two data flows as the
output ones, as in the ``Spectrum'' mode, but now four: the power
spectra of each channel, and also real (cospectrum) and imaginary
(quadrature spectrum) parts of a cross-spectrum. These may allow
finding the coherence function which value varies from 0
(absolutely different signals) to 1 (signals are identical). That
is the reason why the data flow rate increases twice in this case.
During summer observations of 2009, and partially in 2010, the
heliograph was employed first in the ``Spectrum'' mode. The DSPZ
inputs were fed with the sum-difference signal. However such a
data recording method is prone to appreciable loss of information
on the power spectra in separate antennas and signal phase.
Thereby, at summer-end of 2010, the heliograph configuration was
changed, and the signal after switchboards of N-S and E-W arrays
entered the corresponding DSPZ channels, without the
sum-difference unit. In such a situation it is expedient to record
the data using the ``Correlation'' mode due to its aforesaid
advantages and possibility of frequency channels averaging in case
of necessity.

So, in the present heliograph configuration with sequential
formation and recording of picture elements, the two ADC inputs of
the DSPZ unit are fed with the outputs of N-S and E-W arrays. The
N-S array has five outputs which correspond to five beams. These
outputs are sequentially connected to the receiver-recorder (DSPZ)
input. In the end of each cycle of beam inquiry, the auxiliary
noise generator is connected in place of antenna to the both
inputs of the receiver-recorder to form the separating tracks
(markers) on heliograms.

The heliograph receiver-recorder functions as an amplifier and
filter of received signals, shapes the array pencil-beam pattern
and performs raster display in the scan region.

\section{Observation Examples}\label{par11}
In the course of the last decade, the heliographic observations
with the UTR-2 array were largely made occasionally, rather for
testing this instrument subjected to upgrading. The observation
results have allowed to ascertain the first priority in
improvement of the hardware and the software heliograph parts.
Thereto, the indicated period has observed a high solar activity, and
the principal problem in solar observations was radiospectroscopic
research of solar burst activity at decameter wavelengths (see
e.g. \cite{melnik04,chernov,konov07}).

\begin{figure*}
\centering
\resizebox{1.\columnwidth}{!}{%
  \includegraphics[width=1.\textwidth]{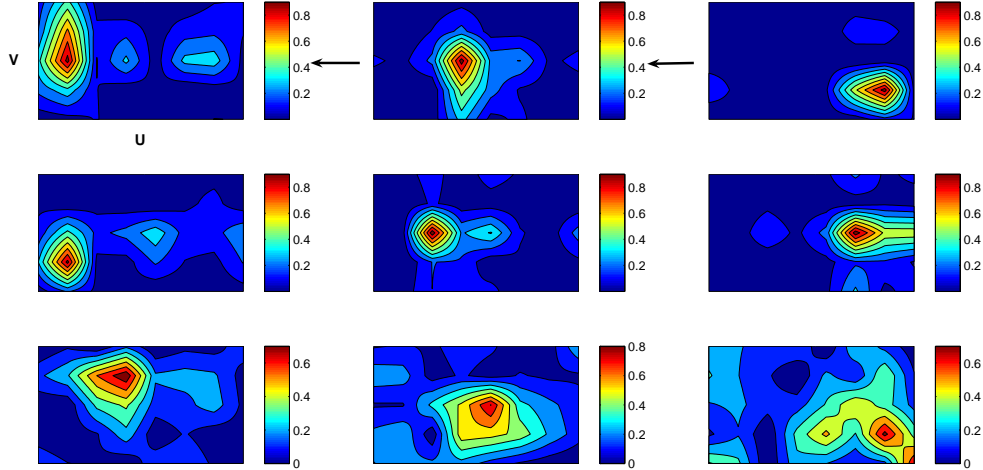}
} \caption{``Raw'' (dirty) maps of sources 3C123 (upper row), 3C348
(middle) and solar corona (lower row) at $\sim$ 21 MHz. Three
images show the sources motion observed owing to the celestial
sphere diurnal rotation (from right to left). The images carry the
UTR-2 sidelobes contribution. The radiation intensity is
normalized to unit, the background is subtracted (i.\ e. being
zero). The dark red color corresponds to maximum intensity, while
the dark blue to minimum.} \label{fig9}
\end{figure*}

Summarizing of the upgrading can be referred to observations in
summer of 2010, which results we are going to present in this
Section. These researches have included night observations of sky
radio sources 3C144 (Crab Nebula), 3C274 (Virgo А), 3C405 (Cygnus
А), etc. For our heliograph, they are considered as point sources
of radio emission. Besides, they have different intensity at
decameter wavelengths and may permit to estimate the heliograph
sensitivity as a whole. The source tracking rate was taken equal
to 16 min. For this time, the source traversed the heliogram frame
from its one edge to the other. The switching rate of the fast
scan phase shifter was taken such that the heliograph image was
formed for 2 min. This is quite enough for the considered case,
when the changes in source position are largely caused by the
celestial sphere's diurnal motion. Such an image formation rate
and source tracking time have permitted us to observe the source
passage from one beam of a frame to the neighbor beam in the next
frame, thus showing the source crossing the whole frame (see
Fig.~\ref{fig9}, upper and middle rows), consisting, as we
remember, of 5-beam eight positions. In the DSPZ ``Correlation''
mode, for each position in the heliograph frame we have succeeded
in formation the pencil-beam pattern.

After successful observations of the aforesaid radio sources, the
similar mode was applied for the quiet Sun observations, too.
These latter were made in the late August and early September,
2010. For the decameter range heliograph, the solar corona appears
to be an extended source of radio emission. Therefore on the
heliogram, the source will be ``distributed'' over several pixels.
Remind that the whole image of the heliograph scan region consists
of   elements (or pixels). One of the major tasks implied by the
heliograph application consists in observations of 2D
distributions of the undisturbed Sun's corona brightness. As is
known, the 11-year solar minimum occurred in 2008, therefore the
quiet Sun power flux densities can be measured in the years near
the solar minimum. The examples of solar corona images obtained
during observations from August 28 till September 3, 2010 are
shown in the lower row of images in Fig.~\ref{fig9}, which were
recorded on August 29 at $\sim$ 7:20 UT. Here, for now, shown are
merely the ``raw'' (or dirty) maps (due to the UTR-2 pattern sidelobes
contribution), but they clearly reveal the corona (or its
brightest part) transition across the heliograph frame due to the
celestial sphere's diurnal motion, that suggests a sufficient
instrument sensitivity even for a single frequency channel. This
important result highly encourages and stimulates further
observations of the solar corona with the UTR-2 array based
heliograph.

The here-shown presentation of observations made with the
heliograph has only a demonstrational (preliminary) character. The
formation of solar corona images using the data of the said
observations at decameter wavelengths with their calibration,
``cleaning'' of images and comparison with the solar events
occurred, anticipate our further efforts to end up being a
separate paper.

\section{Conclusions}\label{par12}
This paper considers a functional scheme of the decameter
wavelength heliograph. The heliograph is based on the UTR-2
antenna system. Any radio source is slowly tracked by the UTR-2
radio telescope standard hardware facility. Within the heliograph
field of view, this antenna pattern beam is guided by a specially
designed phase shifter. The heliograph is recording the image with
the sequential signal injection in each position of the pencil
beam of the UTR-2 array out of its 40 positions available in a
separate frame of the heliograph. The heliograph image repetition
rate can be taken both rather large (suitable for analyzing the
spatial characteristics of fast solar bursts), and a small one
(convenient for the quiet Sun upper corona research).

The image repetition rate, owing to the motion of a five-beam set
of the UTR-2 array within the heliograph scanned area, should be
much greater than the Sun tracking rate. The UTR-2 phasing system
cannot ensure the beam position change with such a speed since it
uses the electromagnetic relays as the switched elements. That is
why the UTR-2 phasing system is used only for the Sun position
tracking. For fast scanning the scan sector, the UTR-2 phasing
system comprises additional phase shifters possessing a fairly
high operating speed and long lifetime. As the scanning is
performed within small angles, the optimum alternative will be
enabling the additional phase shifters between the section outputs
and appropriate inputs of the E-W antenna sections phasing system.
In this case, at fast scanning, the section pattern position
remains the same, and the array beam moves within the section beam
pattern width owing to the additional phase shifters. This results
in lower received-signal intensity of the source situated near the
scan sector edge due to signal strength depression on the section
pattern slopes. In recorded-data processing this fact should be
taken into account, and the respective corrections made in
processing the observation results.

The design features of the decameter wavelength heliograph based
on the UTR-2 array are such that we may belong it to the unique
radio astronomy instruments. Here, suffice it to mention that
among all heliographs used world-wide, this one operates the
lowest frequencies for the study of cosmic radio emission.
Secondly, it is superbroad-band covering practically the entire
range within 10 to 30 MHz. Thirdly, its sensitivity, time and
frequency resolutions, dynamic range, signal space selection are
record-breaking for the given frequency range, and consequently
the problems which can be solved with this instrument have no
world counterparts. The heliograph is also unique in its
capability of obtaining quiet Sun's corona images for about 1 s
period at distances from two to three solar radii, counted from
the Sun's center, and practically without skipping in the $(U,
V)$-plane. The heliograph's sensitivity allows observation of
space-time evolution of low-contrast formations (fine structure of
radio bursts, fibrous structure of Type II bursts, etc.). Besides,
its dynamic range makes it possible to observe the details
simultaneously with the bright radio events (strong Type III
bursts, or against the background of Type II and IV bursts). The
instrument has undergone a long-continued development, and in
retrospect we may think with absolute certainty that the
state-of-the-art heliograph performances have justified the
expended energies.

\section{Acknowledgements}
The authors are very grateful to V. V. Zakharenko for his useful
remarks and discussion of our results, to S. V. Stepkin and V. L.
Kolyadin for the useful pieces of advice, to V. V. Dorovskyy for
the help in accomplishment of this work.

\end{document}